\documentclass[11pt]{article}
\usepackage{epsfig,sint,cite,amsfonts}

\usepackage[english]{babel}
\usepackage{simplewick}

\newcommand{\be}{\begin{equation}}
\newcommand{\ee}{\end{equation}}
\newcommand{\bea}{\begin{eqnarray}}
\newcommand{\eea}{\end{eqnarray}}
\newcommand{\id}{1\!\!1}
\newcommand{\Pg}{\hat{{\rm P}}_{\rm G}}
\newcommand{\Pmu}{\hat{\rm P}^{(\mu)}}

\newcommand{\Tr}{\,\hbox{\rm Tr}}
\newcommand{\R}{\hbox{\rm \tiny R}}
\newcommand{\Rb}{\hbox{\rm R}}
\newcommand{\calR}{{\cal R}}
\newcommand{\nub}{\mbox{\boldmath$\nu$}}
\newcommand{\nubs}{{\mbox{\scriptsize \boldmath$\nu$}}}
\newcommand{\Ps}{\mbox{\tiny ${\cal P}$}}
\newcommand{\Cs}{\mbox{\tiny ${\cal C}$}}
\newcommand{\As}{\mbox{\tiny ${\rm A}_1$}}

\begin{document}

\begin{titlepage}
\begin{flushright}
\hfill CERN-PH-TH/2010-197\\
\hfill HIM-2010-02\\
\hfill MKPH-T-10-40
\end{flushright}

\vskip 2.5 cm
\begin{center}

{\Large\bf A novel approach for computing glueball masses and\\[0.175cm] 
           matrix elements in Yang-Mills theories on the lattice\\[0.5ex]}

\end{center}
\vskip 0.5 cm
\begin{center}
{\large Michele Della Morte$^{\scriptscriptstyle a}$ 
and Leonardo Giusti$^{\scriptscriptstyle b}$}
\vskip 0.75cm
$^{\scriptstyle a}$ Institut f\"ur Kernphysik and Helmholtz Institut,
 University of Mainz,\\
Johann-Joachim-Becher Weg 45, D-55099 Mainz, Germany\\
\vskip 1.5ex
$^{\scriptstyle b}$ CERN, Physics Department, 1211 Geneva 23, Switzerland and \\
     Dipartimento di Fisica, Universit\'a di Milano-Bicocca,\\ 
     Piazza della Scienza 3, I-20126 Milano, Italy\\
\vskip 2.0cm
{\bf Abstract}
\vskip 0.35ex
\end{center}

\noindent
We make use of the global symmetries of the Yang-Mills theory on
the lattice to design a new computational strategy for extracting 
glueball masses and matrix elements which achieves an exponential
reduction of the statistical error with respect to standard 
techniques. By generalizing our previous work on the parity symmetry,
the partition function of the theory is decomposed into a sum of path 
integrals each giving the contribution from multiplets of 
states with fixed quantum numbers associated to parity, charge conjugation, 
translations, rotations and central conjugations $Z_N^3$. Ratios of path 
integrals and correlation functions can then be computed with a multi-level 
Monte Carlo integration scheme whose numerical cost, at a fixed statistical 
precision and at asymptotically large times, increases power-like with the 
time extent of the lattice. The strategy is implemented for the SU(3) 
Yang--Mills theory, and a full-fledged computation of the mass and 
multiplicity of the lightest glueball with vacuum quantum numbers is 
carried out at a lattice spacing of 0.17~fm.
\vfill

\eject

\end{titlepage}

\section{Introduction}
The existence of glueballs is a distinctive property of quantized 
non-Abelian gauge theories~\cite{Fritzsch:1972jv,Fritzsch:1975tx,Coleman:1977hd}. 
Since the first Monte Carlo simulations of lattice field theories, 
glueballs have been the focus of many studies (see Ref.~\cite{Berg:1982sd} 
and reference therein). The main difficulty for isolating their contribution
in correlation functions, and thus computing their masses and 
matrix elements, was identified almost immediately: 
the signal-to-noise ratio of suitable two-point correlation functions~\cite{Berg:1982kp} 
decreases exponentially with the time separation of the sources, and in practice 
it is very difficult to find a window where statistical and systematic 
errors are both under control~\cite{Parisi:1983ae,Lepage:1989hd}. 
A widely used strategy to mitigate this problem is to reduce 
the contamination from excited states in the correlators by constructing 
interpolating fields with a small overlap over them~\cite{Albanese:1987ds,Teper:1987wt}. 
The lowest energy is then extracted at short time-distances by assuming a 
negligible contribution from excited states, sometimes also with the help of anisotropic
lattices~\cite{Morningstar:1997ff}. The most comprehensive studies of glueball masses
and matrix elements performed with these techniques can be found in 
Refs.~\cite{Albanese:1987ds,Bali:1993fb,Vaccarino:1999ku,
Morningstar:1999rf,Chen:2005mg,Meyer:2008tr}.
This approach is not entirely satisfactory from a conceptual and a practical 
point of view. The problem of the exponential degradation of the signal-to-noise ratio remains 
unsolved, and the functional form of the sources is usually optimized so that 
correlators or combinations of them show a single exponential decay in the 
short time-range allowed by the statistical accuracy. A solid evidence that a single state 
dominates, i.e.  a long exponential decay over many orders of magnitude, is thus missing. 
In most of the cases the computation of more involved correlation functions 
remains inaccessible. 

Recently we proposed a new approach to solve the 
problem~\cite{DellaMorte:2008jd,DellaMorte:2007zz} 
(see also \cite{DellaMorte:2009rf,DellaMorte:2010}). 
By using the transfer matrix formalism, it is possible to introduce in the partition function 
projectors which select the contributions from states with a given set of quantum numbers only. 
The composition rules of the corresponding transfer matrix elements can then be exploited 
to implement a hierarchical 
multi-level integration procedure. By iterating over several levels, the numerical cost for 
computing ratios of partition functions or correlation functions grows, at asymptotically 
large times, with a power of the time extent of the lattice rather than 
exponentially.

The aim of this paper is to design a new strategy for computing the lightest 
glueball masses, multiplicities and their matrix elements in each super-selection sector of the 
Hilbert space of the Yang-Mills theory. This is achieved by generalizing the 
analysis in Ref.~\cite{DellaMorte:2008jd} to all discrete symmetries of the 
lattice theory in finite volume: parity, charge conjugation, rotations, translations 
and central conjugations $Z_N^3$. We assume some familiarity with 
Ref.~\cite{DellaMorte:2008jd}, especially in Section~6, where the numerical 
algorithm is described. Being based on the transfer matrix formalism, 
the approach is rather general and can be applied to a wide class of bosonic 
theories. 

We implement the strategy in the SU(3) Yang--Mills theory on lattices with a spacing of 
roughly $0.17$~fm, spatial volumes up to $5\, \mbox{fm}^3$, and time extent up to 
$2\, \mbox{fm}$. A full-fledged computation of the mass and multiplicity of the lightest 
glueball with vacuum quantum numbers is then carried out. The algorithm behaves as expected, 
and the multi-level integration scheme achieves an exponential enhancement of the 
signal-to-noise ratio for the quantities considered. As a result we can 
extract with confidence the contribution from the lightest glueball 
and, from a lattice with a time extent of $2$~fm, determine its mass with a 
precision of few percent. Its multiplicity turns out to be one within statistical 
errors, and the latter are small enough to exclude all other values 
allowed by the underlying group theory.

\section{Preliminaries and basic notation}
For definiteness we focus on the SU(3) Yang--Mills theory 
discretized by the standard plaquette Wilson action. The theoretical 
discussion, however, can be applied to other bosonic field theories,
and it is mostly independent on the discretization details.
The theory is set up on a finite four-dimensional lattice of volume 
$V=T\times L^3$ with a spacing $a$ and periodic boundary 
conditions\footnote{Throughout the paper 
dimensionful quantities are always expressed in units of $a$ unless 
explicitly specified.}. 
The gauge-invariant action is defined as (all unexplained notation can be found 
in Ref.~\cite{DellaMorte:2008jd})
\be
S[U] = \frac{\beta}{2}\, \sum_{x} \sum_{\mu,\nu} 
\left[1 - \frac{1}{3}{\rm Re}\Tr\Big\{U_{\mu\nu}(x)\Big\}\right]\; ,
\ee
where the trace is over the color index, and $\beta=6/g_0^2$ with $g_0$ 
being the bare coupling constant. The plaquette is defined as a function 
of the gauge links $U_\mu(x)$ as
\be\label{eq:placst}
U_{\mu\nu}(x) = U_\mu(x)\, U_\nu(x+\hat \mu)\, U^\dagger_\mu(x + \hat \nu)\,
             U^\dagger_\nu(x)\; ,  
\ee
with $\mu,\,\nu=0,\dots,3$, $\hat \mu$ is the unit vector along the 
direction $\mu$, and $x$ is the space-time coordinate. The path integral 
is defined as usual
\be\label{eq:Zstand}
Z = \int {\rm D}_4 [U] \; e^{-S[U]}\; ,\qquad 
{\rm D}_4 [U] = \prod_{x}\, \prod_{\mu=0}^{3} {\cal D} U_\mu(x)\; ,
\ee
where ${\cal D} U$ is the invariant Haar measure on the SU(3) group, which 
throughout the paper will be always normalized so 
that $\int {\cal D} U = 1$. The ``coordinate'' basis in the Hilbert space 
of the theory is the set of vectors which diagonalize the field operators at 
all spatial points, so that on a given time-slice of the lattice 
\be
\hat {\rm U}_k({\bf x}) |U_{x_0}\rangle = U_k(x_0,{\bf x}) |U_{x_0}\rangle\; .
\ee
As the notation suggests, the operator eigenvalues $U_k(x_0,{\bf x})$ are 
identified with the spatial links on the time-slice $x_0$ at the spatial 
coordinate ${\bf x}$. The matrix elements of the transfer 
operator $\hat {\rm T}$ are 
\be
{\rm T}\Big[U_{x_0+1},U_{x_0}\Big] = 
\left\langle U_{x_0+1}| \hat {\rm T}\Pg | U_{x_0} \right\rangle\; , 
\ee
where $\Pg$ is the projector onto the gauge invariant states. In the following, 
for notational simplicity, this projector is always included in the 
definition of $\hat {\rm T}$. 
For the Wilson action the transfer matrix elements 
are~\cite{Wilson:1977nj,Luscher:1976ms,Creutz:1976ch,Osterwalder:1977pc}
\be\label{eq:TW}
{\rm T}\Big[U_{x_0+1},U_{x_0}\Big] = \int {\rm \bf D}[\Omega]
\; e^{-L[U^{\Omega}_{x_0+1},U_{x_0}]} \; ,
\qquad
{\rm \bf D}[\Omega] = 
\prod_{{\bf x}} {\cal D}\Omega({\bf x})\; , 
\ee
where the explicit form of the Lagrangian $L$ can be found in 
Appendix \ref{app:appA}. From this formula one 
can define the transfer matrix elements associated to a given thick 
time-slice, i.e. the ensemble of points in the sub-lattice with time
coordinates in a given interval $[x_0,y_0]$ and bounded by the equal-time
hyper-planes at times $x_0$ and $y_0$, as 
\be\label{eq:T1}
{\rm T}\Big[U_{y_0},U_{x_0}\Big] = \int \prod_{w_0=x_0+1}^{y_0-1}\, 
{\rm \bf D}_3[U_{w_0}]\, 
\prod_{z_0=x_0}^{y_0-1} {\rm T}\Big[U_{z_0+1},U_{z_0}\Big]\;,
\qquad
{\rm \bf D}_3[V] = \prod_{{\bf x}}\, \prod_{k=1}^{3} 
{\cal D} U_k({\bf x})\; .
\ee
By identifying the gauge transformations $\Omega$ in Eq.~(\ref{eq:TW}) 
with the links in the temporal direction, the path integral can
thus be written as 
\be\label{eq:Z1}
Z = \int \prod_{x_0=0}^{T-1}\, 
{\rm \bf D}_3[U_{x_0}]\, {\rm T}\Big[U_{x_0+1},U_{x_0}\Big]\; ,
\ee
which corresponds to 
\be\label{eq:Z2}
Z = \Tr \left\{\hat {\rm T}^T \right\}\; 
\ee 
with the trace being over all gauge invariant states. 

\section{Decomposition of path integrals\label{sec:dec}}
The partition function $Z$ can be decomposed into a 
sum of path integrals each giving the contribution from multiplets of 
states which transform as an irreducible representation of a symmetry 
group of the theory. The phase space of the theory can indeed be divided 
into regular representations of the group, which in turn can be 
decomposed into irreducible ones by applying the standard group-theory 
machinery~\cite{Cornwell:1984co,Hamermesh:1989ha}. The invariance of the 
transfer operator $\hat {\rm T}$ under the group transformations then guarantees 
the decomposition of the partition function. An analogous decomposition
applies to the correlation functions of the theory.

To carry out this program in detail, let us assume that the theory is 
invariant under the 
transformations of a generic discrete\footnote{The generalization to 
Lie groups is straightforward.} symmetry group ${\cal G}$ of order $g$. Its 
elements $\calR_i$, with $i=1,\dots,g$, act on a generic vector 
$| U \rangle$  of the coordinate basis of the Hilbert space as
\be
\hat \Gamma(\calR_i) |U\rangle = |U^{\calR_i}\rangle\; ,
\ee 
where $U^{\calR_i}$ is the gauge field obtained by applying the 
group transformation $\calR_i$ to the original one (see next section). 
The vectors $|U^{\calR_i}\rangle$ form a regular representation, i.e.  
\be
\hat \Gamma(\calR_i) |U^{\calR_j}\rangle = \sum_{l=1}^{g}
\Gamma^{\rm reg}_{lj}(\calR_i)|U^{\calR_l}\rangle\; ,
\ee
where 
\be
\Gamma^{\rm reg}_{lj}(\calR_i) = 
\left\{\begin{array}{cc}
1 & \mbox{if} \;\;\; \calR_l=\calR_i \calR_j\\[0.5cm]
0 & \mbox{if} \;\;\; \calR_l\neq \calR_i \calR_j
\end{array}\right.\; .
\ee
The non-equivalent irreducible representations  $\Gamma^{(\mu)}(\calR_i)$
are labeled by $\mu=1,\dots,N_r$, where  $N_r$ is the number of classes 
in which the group can be divided, their dimensions $n_\mu$ satisfy 
\be
\sum_{\mu=1}^{N_r} n^2_\mu = g\; ,
\ee
and their characters are given by
\be
\chi^{(\mu)}(\calR_i)=\sum_{j=1}^{n_\mu}\Gamma^{(\mu)}_{jj}(\calR_i)\; .
\ee
For a given realization of an irreducible representation $\mu$,
the ``projector'' operators are defined as usual by
\be\label{eq:Projl}
\hat{\rm P}^{(\mu)}_{jl}= \frac{n_\mu}{g} \sum_{i=1}^{g}\, 
\Gamma^{(\mu)\, *}_{jl}(\calR_i)\,\hat \Gamma(\calR_i)\, ,
\ee
and they satisfy 
\be
\hat{\rm P}^{(\mu)\,\dagger}_{jl} = \hat{\rm P}^{(\mu)}_{lj}\; , \qquad
\hat{\rm P}^{(\mu)}_{jl}\, \hat{\rm P}^{(\nu)}_{mn} = 
\delta^{\mu\nu}\, \delta_{lm}\, \hat{\rm P}^{(\mu)}_{jn}\; ,
\qquad
\sum_{\mu=1}^{N_r}\sum_{j=1}^{n_\mu} \hat{\rm P}^{(\mu)}_{jj} = \hat \id \; .
\ee
In particular $\hat{\rm P}^{(\mu)}_{jj}$ is the projector onto states of 
the Hilbert space that transform as the component $j$ of an irreducible 
representation $\mu$. The path integral can then be decomposed as 
\be\label{eq:zmuh}
Z = \sum_{\mu=1}^{N_r} Z^{(\mu)}\; , \qquad 
Z^{(\mu)} = \sum_{j=1}^{n_\mu} \Tr \left\{\hat {\rm T}^T \hat{\rm P}_{jj}^{(\mu)} 
\right\} \; .
\ee
By inserting a complete set of eigenstates of the Hamiltonian 
we can also write 
\be\label{eq:weigts}
Z^{(1)} = e^{-E_0\, T}  \left[1 + \sum_{m} w^{(1)}_m\, e^{-E^{(1)}_m T}\right]\, , 
\qquad  Z^{(\mu)} = e^{-E_0\, T} \sum_{m} w^{(\mu)}_m\, e^{-E_m^{(\mu)} T}\; , 
\ee
where $\mu=1$ corresponds to the invariant singlet representation. In these 
expressions $E_0$ is the vacuum energy, $E_m^{(\mu)}$ 
are the energies (with respect to the vacuum one) of eigenstates 
in the sector $\mu$, and $w^{(\mu)}_m$ are the corresponding multiplicities. 
The latter are integers and positive since for the Wilson action the transfer 
operator $\hat {\rm T}$ is self-adjoint and strictly 
positive~\cite{Luscher:1976ms}.

\subsection{Partition function}
The transfer matrix elements among states belonging to 
irreducible representations can be written as
\be
{\rm T}^{(\mu)}_{jj} \Big[U_{x_0+1},U_{x_0}\Big] = 
\frac{n_\mu}{g} \sum_{i=1}^{g}\, \Gamma^{(\mu)}_{jj}(\calR_i)\,
\left\langle U^{\calR_i}_{x_0+1}|\, \hat {\rm T}\, | U_{x_0} \right\rangle\; .
\ee  
For a thick time-slice the analogous ones are defined  
by exploiting the orthogonality of the projectors 
\be\label{eq:compmu}
\delta^{\mu\nu}\,\delta_{jl}\, {\rm T}^{(\mu)}_{jj}\Big[U_{y_0},U_{x_0}\Big] = 
\int {\rm \bf D}_3[U_{z_0}]\,{\rm T}^{(\mu)}_{jj}\Big[U_{y_0},U_{z_0}\Big]\;
{\rm T}^{(\nu)}_{ll}\Big[U_{z_0},U_{x_0}\Big]\; , 
\ee
which also implies 
\be\label{eq:thick}\displaystyle
 \frac{{\rm T}_{jj}^{(\mu)}[U_{y_0},U_{x_0}]}
      {{\rm T}\;[U_{y_0},U_{x_0}]}=
\frac{1}{{\rm T}\;[U_{y_0},U_{x_0}]}\int
 {\rm D}_4 [U] \; e^{-S[U]}\,
\frac{{\rm T}_{jj}^{(\mu)}[U_{y_0},U_{y_0-1}]}
     {{\rm T}\,[U_{y_0},U_{y_0-1}]}\; ,
\ee
\noindent where the integration is restricted to the active links of the 
thick time-slice. The relations~(\ref{eq:compmu}) and (\ref{eq:thick})
are the basic building blocks for the practical implementation of the 
multi-level integration algorithm described in Section \ref{sec:algo}. 
Finally, by repeatedly using  Eq.~(\ref{eq:compmu}), it is possible to 
rewrite the path integral for a given representation as 
\be\label{eq:Zmu}
 Z^{(\mu)} = \sum_{j=1}^{n_\mu} \int \prod_{x_0=0}^{T-1} {\rm \bf D}_3[U_{x_0}]\,
{\rm T}_{jj}^{(\mu)} \Big[U_{x_0+1},U_{x_0}\Big]  \; .
\ee
It is interesting to notice that even though the transfer matrix formalism 
inspired the construction, the above considerations hold independently of the 
existence of a positive self-adjoint transfer operator. The insertion of 
${\rm T}_{jj}^{(\mu)} \Big[U_{x_0+1},U_{x_0}\Big]$ 
in the path integral plays the same r\^ole as $\Pmu_{jj}$ in Eq.~(\ref{eq:zmuh}),
i.e. it allows the propagation in the time direction of states 
belonging to irreducible representations $\mu$ only (component $j$). 

\subsection{Correlators of composite operators}
We are interested in irreducible tensor operators, which under the group 
transform as 
\be\label{eq:rotO}
\hat \Gamma(\calR_i)\,  \hat O^{(\mu)}_{j}\, 
\hat \Gamma^\dagger(\calR_i) =  
\sum_{l=1}^{n_\mu}\Gamma^{(\mu)}_{l j}(\calR_i)\, \hat O^{(\mu)}_{l}\, ,
\ee
that are diagonal in coordinate space, i.e. 
\be
\hat O^{(\mu)}_{j} \left| U \right\rangle = 
O^{(\mu)}_{j} (U) \left| U \right\rangle
\ee
where $O^{(\mu)}_{j} (U)$ is a functional of the links. The 
transformation rule (\ref{eq:rotO}) implies that 
\be
O^{(\mu)}_{j} (U^{\calR_i}) = \sum_{l=1}^{n_\mu}\Gamma^{(\mu)\, *}_{jl}(\calR_i)\, 
O^{(\mu)}_{l}(U)\; .
\ee
By using the Wigner-Eckart theorem, we can define the reduced matrix 
elements corresponding to an operator insertion on the time-slice $x_0$
inside a thick time-slice of size $d$ as
\bea
\displaystyle \Big[{\cal O}^{(\mu)}_k(x_0)\Big]^{(\mu_1)(\mu_2)}_{j_1\,j_2}
\Big[U_{z_0+d},U_{z_0}\Big] & = &  \displaystyle
\frac{1}{n_{\mu_1}}\sum_{l=1}^{n_\mu} \sum_{l_1=1}^{n_{\mu_1}} 
\sum_{l_2=1}^{n_{\mu_2}}
\left(^{\mu_2}_{l_2}\, ^\mu_{l_{}}\Big|\, ^{\mu_1}_{l_1}\, ^k\right) \times\\[0.25cm]
& & \displaystyle
\left\langle U_{z_0+d}\right|\hat{\rm P}^{(\mu_1)}_{j_1 l_1} 
\hat {\rm T}^{(z_0+d-x_0)} \hat O^{(\mu)}_{l} \hat {\rm T}^{(x_0-z_0)}
\hat{\rm P}^{(\mu_2)}_{l_2 j_2} \left| U_{z_0} \right\rangle\; , \nonumber
\eea
where $z_0\leq x_0\leq z_0+d$, 
$\left(^{\mu_2}_{l_2}\, ^\mu_{l_{}}\Big|\, ^{\mu_1}_{l_1}\, ^k\right)$
are the Clebsch--Gordan coefficients, $k=1,\dots,n_{\mu_1}^{\mu \mu_2}$ and 
$n_{\mu_1}^{\mu \mu_2}$ is the number of times that the irreducible representation
$\mu_1$ appears in the direct product representation 
$\Gamma^{(\mu)} \otimes \Gamma^{(\mu_2)}$. Thanks to the orthogonality 
properties of the Clebsch--Gordan coefficients, a two-point correlation function 
can be written as ($w_0\leq y_0\leq w_0+d$)
\bea
& & \sum_{j=1}^{n_\mu} \left\langle O^{(\mu)}_j(y_0) 
O^{(\mu)*}_j (x_0)  \right\rangle =   {{1}\over{Z}}\,
\int {\rm \bf D}_3[U_{w_0+d}]\,{\rm \bf D}_3[U_{w_0}]\, 
{\rm \bf D}_3[U_{z_0+d}]\, {\rm \bf D}_3[U_{z_0}]\, {\rm \bf D}_3[U_{0}]\times\nonumber\\
& \times & \sum_{\mu_1,\mu_2} \sum_{j_1,j_2}
\frac{1}{n_{\mu_2}} \sum_{k=1}^{n_{\mu_1}^{\mu \mu_2}}
{\rm T}^{(\mu_1)}_{j_1 j_1} \Big[U_{0},U_{w_0+d}\Big]
\Big[{\cal O}^{(\mu)}_k(y_0)\Big]^{(\mu_1)(\mu_2)}_{j_1 j_2}\Big[U_{w_0+d},U_{w_0}\Big]
\times\label{eq:factOO}\\[0.25cm]
& \times &
{\rm T}^{(\mu_2)}_{j_2 j_2}\Big[U_{w_0},U_{z_0+d}\Big]
\Big[{\cal O}^{(\mu)}_k(x_0)\Big]^{(\mu_1)(\mu_2)\, *}_{j_1 j_2}\Big[U_{z_0},U_{z_0+d}\Big] 
{\rm T}^{(\mu_1)}_{j_1 j_1} \Big[U_{z_0},U_{0}\Big]\; ,
\nonumber
\eea
where $z_0+d<w_0$. The generalization to $n$-point correlation functions 
is straightforward.

\section{Discrete symmetries of the Yang--Mills theory\label{sec:symm}}
The SU(3) Yang--Mills theory on a lattice of finite volume with periodic 
boundary conditions is invariant under parity, charge conjugation, 
translations, rotations and central conjugations $Z^3_3$. In this 
section we set the notation for these groups, and we briefly review the 
properties which are relevant for the paper.

\subsection{Parity}
The group is of order 2. A regular representation is two-dimensional,
and it is spanned by the vectors
\be
|U^{\calR_1}\rangle = |U\rangle\; , \qquad  
|U^{\calR_2}\rangle = |U^\wp\rangle\; ,
\ee
where on a generic time-slice $x_0$
\be
U^\wp_k(x_0,{\bf x}) = U^\dagger_k(x_0,-{\bf x}-\hat k)\; .
\ee
The two irreducible representations of dimension one are 
$\Gamma^{(\pm)}(\calR_1)=\pm\Gamma^{(\pm)}(\calR_2)=1$, 
where the phase convention is the same as in 
Ref.~\cite{DellaMorte:2008jd}.

\subsection{Charge conjugation}
The group is of order 2. A regular representation is two-dimensional
and it is spanned by
\be
|U^{\calR_1}\rangle = |U\rangle\; , \qquad  
|U^{\calR_2}\rangle = |U^{\cal C}\rangle\; ,
\ee
where 
\be
U^{\cal C}_k(x_0,{\bf x}) = U^*_k(x_0,{\bf x})\; .
\ee
The two irreducible representations of dimension 
one are $\Gamma^{(\pm)}(\calR_1)=\pm\Gamma^{(\pm)}(\calR_2)=1$.

\subsection{Translations}
The group of translations is a direct product of three Abelian groups, one
for each space direction. Its elements are labeled
by a three dimensional vector of integers ${\bf m} = ({m_1,m_2,m_3})$,
with ${m_i}=0,\dots,L-1$, where each component labels the elements
of the Abelian group in the corresponding direction. A regular 
representation is $L^3$-dimensional and is spanned by
\be
|U^{\calR_{\bf m}}\rangle = |U^{\bf m}\rangle\
\ee
where 
\be
U^{\bf m}_k(x_0,{\bf x}) = U_k(x_0,{\bf x}-{\bf m})\; .
\ee
Since the group is Abelian, each element forms its own class
and there are $L^3$ inequivalent irreducible representations 
of dimension 1 
\be
\Gamma^{(\bf p)}(\calR_{\bf m}) = e^{i\, {\bf p \cdot m}}\; 
\ee 
which are labeled
by momentum vectors $\displaystyle {\bf p}=\frac{2\pi}{L}\, [n_1,n_2,n_3]$, with
$n_i=0,\dots,L-1$. 

\subsection{Rotations}
The octahedral group is of order 24. Its elements are listed in 
Appendix \ref{app:appB}.  They form 5 equivalence classes. 
A regular representation is 24-dimensional and is spanned by 
\be
|U^{\calR_i}\rangle = |U^{\R_i}\rangle\; , \qquad i=1,\dots,24
\ee
where 
\be
U^{\R_i}_h (x_0,{\bf x}) = 
\left\{\begin{array}{ll}
U_k(x_0,\Rb^{-1}_i{\bf x})  & (\Rb_i)_{hk}>0 \\[0.5cm]
U^\dagger_k(x_0,\Rb^{-1}_i{\bf x} - \hat k )  & (\Rb_i)_{hk}<0
\end{array}\right.\; .
\ee
The inequivalent irreducible representations are two singlets 
${\rm A}_1$ and ${\rm A}_2$, one doublet ${\rm E}$ and two 
triplets ${\rm T}_1$ and ${\rm T}_2$. Their expressions and their
characters are given in Appendix \ref{app:appB} while the Clebsch--Gordan 
coefficients can be found in Ref.~\cite{Cornwell:1984co,Hamermesh:1989ha}.

\subsection{Central conjugations ${\bf Z^3_3}$}
The presence of this symmetry, first described by 't Hooft 
in Ref.~\cite{'tHooft:1979uj} (see also Ref.~\cite{Luscher:1988sd} for a review),
is due to the choice of periodic boundary conditions and it disappears 
in the infinite volume limit. The group is a direct product of 
three $Z_3$, one for each spatial direction. It is of order 27, and 
its elements are labeled by a three dimensional vector 
of integers $\nub=(\nu_1,\nu_2,\nu_3)$, with $\nu_i=0,1,2$,
where each component labels the elements of the Abelian 
group in the corresponding direction. A regular representation 
is $27$-dimensional and is spanned by
\be
|U^{\calR_{\nubs}}\rangle = |U^{\nubs}\rangle\
\ee
with 
\be
U^{\nubs}_k(x_0,{\bf x}) = 
\Lambda_{\nubs}({\bf x})\, U_k(x_0,{\bf x})\,
\Lambda^\dagger_{\nubs}({\bf x} + \hat k)\; ,
\ee
and 
\be
\Lambda_{\nubs}({\bf x}) = e^{\,i \frac{2\pi}{3L}({\nubs}\cdot{\bf x})\, W}\;, 
\qquad \Lambda_{\nubs}({\bf x}+L\,\hat{k}) = e^{i \frac{2\pi}{3}\,\nu_k}  
\Lambda_{\nubs}({\bf x})
\ee
where $W$ is a $3 \times 3$ diagonal matrix with elements 
$W_{\alpha\alpha}=(1-3\,\delta_{\alpha 3})$.
Since the group is Abelian, each element forms its own class,
and there are $27$ non-equivalent irreducible ones of dimension 1
\be
\Gamma^{(\bf e)}(\calR_\nubs) = e^{i\, {\bf e \cdot} \nubs}\;
\ee
which are labeled by the electric flux vectors 
$\displaystyle {\bf e}=\frac{2\pi}{3}\, [e_1,e_2,e_3]$, 
with $e_i=0,1,2$. 

\section{Glueball masses, multiplicities and matrix elements\label{sec:gmm}}
Dynamical properties of glueballs can be extracted from ratios of partition 
functions and correlators, which in turn can be computed efficiently by 
judiciously putting together the tools developed in the previous sections. 
As usual, the various super-selection sectors of the Hilbert space 
are identified by the quantum numbers associated to a complete set 
of operators which commute among themselves and with the transfer operator.
In the zero-momentum sector $({\bf p}={\bf 0})$, glueball states are classified 
by their transformation properties under rotations $(\mu,j)$, parity 
$({\cal P})$ and charge conjugation $({\cal C})$. Moreover in finite volume a 
null electric flux vector $({\bf e}={\bf 0})$ identifies the ``physical sector'' of 
the theory, i.e. the one which survives in the infinite volume limit. 
The corresponding projectors are given by
\be
\hat{\rm P}_{jj}^{(\mu,\Ps,\Cs)} = 
\hat{\rm P}_{jj}^{(\mu)} \hat{\rm P}^{(\Ps)}
\hat{\rm P}^{(\Cs)} \hat{\rm P}^{({\bf p}=0)} 
\hat{\rm P}^{({\bf e}=0)}\; ,
\ee
and the partition functions read
\be
Z^{(\mu,\Ps,\Cs)} = \sum_{j=1}^{n_\mu}
\Tr \left\{\hat {\rm T}^T \hat{\rm P}_{jj}^{(\mu,\Ps,\Cs)} \right\}\; .
\ee
The transfer matrix elements associated to a given thick-time slice are defined as 
\be\label{eq:prjtut}
{\rm T}^{(\mu,\Ps,\Cs)}_{jj}
\Big[U_{y_0},U_{x_0}\Big] = \frac{L^{-3}}{2592} \sum_{\R_i,p,c,{\bf m},\nubs}
\chi^{(\mu,\Ps,\Cs)}_{j;\R_i,p,c}\,
{\rm T}\Big[U^{\R_i,p,c,{\bf m},\nubs}_{y_0},U_{x_0}\Big]\; ,
\ee
where 
\be
\chi^{(\mu,\Ps,\Cs)}_{j;\R_i,p,c} = n_\mu\, 
{\cal P}^{p+1}\, {\cal C}^{c+1}\, \, \Gamma^{(\mu)}_{jj}(\Rb_i)
\ee
and for each group the sum is on all its elements. The 
corresponding path integrals are finally given by 
\be\label{eq:parttut}
 Z^{(\mu,\Ps,\Cs)} 
= \sum_{j=1}^{n_\mu} \int \prod_{x_0=0}^{T-1} {\rm \bf D}_3[U_{x_0}]\, 
{\rm T}_{jj}^{(\mu,\Ps,\Cs)} \Big[U_{x_0+1},U_{x_0}\Big]\; .
\ee
In each sector, thanks to Eqs.~(\ref{eq:weigts}) and with the 
exception of the vacuum one (see below), the contribution of the 
lightest glueball is the leading exponential in $T$. Its mass 
and multiplicity can thus be 
extracted from the ratio $Z^{(\mu,\Ps,\Cs)}/Z^{({\rm A}_1,+,+)}$ computed at 
large enough values of $T$, where the contamination from heavier states
can be neglected. 

The sector $(\mu,{\cal P},{\cal C})=({\rm A}_1,+,+)$ is special because the vacuum 
contribution dominates the partition function at large $T$. In this case the mass 
and the multiplicity of the lightest glueball can be determined 
by a two step procedure\footnote{This is the simplest example of how the 
procedure described in this paper can be generalized to the computation 
of excited states in each symmetry sector.}. First, one defines the projector 
onto physical states with non-zero momenta and positive charge conjugation 
\be\label{eq:p+}
\hat{\rm P}^{({\bf p},+)} = \hat{\rm P}^{(\Cs=+)} \hat{\rm P}^{({\bf p})} 
\hat{\rm P}^{({\bf e}=0)}\; ,
\ee
where, for instance, ${\bf p}=[2\pi/L,0,0]$. The vacuum does not contribute
to the corresponding partition function $Z^{({\bf p},+)}$ which, at large $T$, 
is dominated by the lightest glueball state even under charge conjugation. 
Its energy and  multiplicity can be extracted from the large $T$ behaviour of 
$Z^{({\bf p},+)}/Z^{({\bf 0},+)}$. The mass can then be determined, 
up to ${\cal O}(a^2)$ effects, by using the continuum dispersion relation. 
In the second step one defines the projector 
\be\label{eq:vort}
\hat{\rm P}^{(\As,+,+)}_{\perp} = 
\hat{\rm P}^{({\bf p}=0)} \hat{\rm P}^{({\bf e}=0)}
\left\{\id - \hat{\rm P}^{(\As,+,+)}\right\}\; ,  
\ee
computes $Z^{(\As,+,+)}_{\perp}$ defined analogously
as in Eq.~(\ref{eq:parttut}), and extract the mass and 
the multiplicity of the lightest state. If, in the continuum limit, 
the mass turns out to be heavier than the one computed in the first step, 
the latter is the mass of the lightest glueball with vacuum quantum 
numbers, and it is also the mass gap of the theory.

By generalizing Eq.~(\ref{eq:factOO}) to the case of multiple quantum
numbers, correlation functions of composite operators can be calculated 
analogously to ratios of partition functions discussed above. 
It is interesting to notice that the approach 
described here allows to determine efficiently also the mass and the 
multiplicity of states in the ``unphysical'' sectors 
(${\bf e} \neq 0$) \cite{Michael:1989leo}.

\section{Numerical algorithm\label{sec:algo}}
\begin{figure}[!t]
\begin{center}
\includegraphics[width=10.0cm]{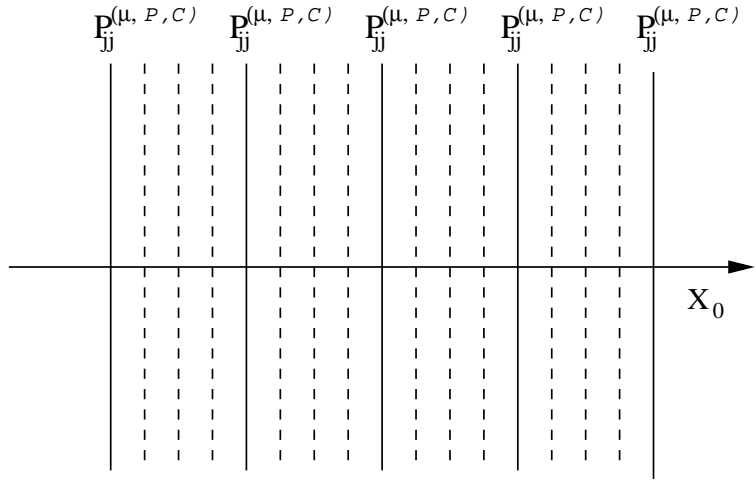}
\caption{Schematic description of the partitioning of the lattice 
in the multi-level integration algorithm.\label{fig:multiL}}
\end{center}
\end{figure}
The formul\ae~in Eqs.~(\ref{eq:prjtut}) and (\ref{eq:parttut}) require 
some manipulation before they can be implemented in a numerical 
simulation. 

\subsection{Thick-time slice with fixed quantum numbers}
In the computation of the transfer matrix elements (normalized to the standard one) given 
in Eq.~(\ref{eq:prjtut}), the basic building blocks are the ratios
\be\label{eq:basicrat}
\frac{{\rm T}\Big[U^{\R_i,p,c,{\bf m},\nubs}_{y_0},U_{x_0}\Big]}
{{\rm T}\Big[U_{y_0},U_{x_0}\Big]}\; .  
\ee
Once written as in Eq.~(\ref{eq:thick}), they are computed numerically
through the telescopic algorithm described in Section 4.1 of 
Ref.~\cite{DellaMorte:2008jd}, the latter being  generalized to all possible 
combinations of group transformations. Their calculation is the most 
expensive part of the multi-level procedure. It is 
therefore worthwhile to optimize on the number of them and/or on their 
numerical precision. To this aim it is relevant 
to notice that a weighted average appears in Eq.~(\ref{eq:prjtut}), 
with the weights given by the appropriate products of characters. 
The larger is the number of significant terms averaged over, 
the lower is the statistical precision required on each term so to achieve 
a given accuracy on the sum. In the case of the singlet 
under all symmetry groups, for instance, the precision on each thick-time 
slice ratio can be reduced proportionally to the square root of the 
number of addenda. A large number of terms therefore 
does not automatically implies a more expensive numerical computation.
The statistical error on each ratio (\ref{eq:basicrat}), however, 
cannot be arbitrarily large for these arguments to apply. Therefore 
it also pays off to reduce the number of ratios to be determined. 
This can be achieved by promoting each group index to a stochastic 
variable with integer values. To each of the index combinations is thus associated a 
probability distribution $\Pi_{\R_i,p,c,{\bf m},\nubs}$ so that
\be
\Pi_{\R_i,p,c,{\bf m},\nubs} \geq 0\;, \quad
\sum_{\R_i,p,c,{\bf m},\nubs} \Pi_{\R_i,p,c,{\bf m},\nubs} = 1\; .
\ee
Moreover if we sum over a subset of indices, for instance $\Rb_i$, then 
\be
\Pi_{p,c,{\bf m},\nubs} = \sum_{\R_i} \Pi_{\R_i,p,c,{\bf m},\nubs}
\ee
is still a probability distribution in the remaining variables. If, for 
instance, we choose $\Pi_{\R_i,p,c,{\bf m},\nubs} = {\rm const}$, then 
the thick-time slice with fixed quantum numbers can be written as 
\be\label{eq:sumrnd}
\frac{{\rm T}_{jj}^{(\mu,\Ps,\Cs)}\Big[U_{y_0},U_{x_0}\Big]}
{{\rm T}\Big[U_{y_0},U_{x_0}\Big]}
= \sum_{\R_i,p,c,{\bf m},\nubs}
\Pi_{\R_i,p,c,{\bf m},\nubs}\; \chi^{(\mu,\Ps,\Cs)}_{j;\R_i,p,c}\;
\frac{{\rm T}\Big[U^{\R_i,p,c,{\bf m},\nubs}_{y_0},U_{x_0}\Big]}
{{\rm T}\Big[U_{y_0},U_{x_0}\Big]}\; . 
\ee
The sum on the r.h.s can be computed stochastically by extracting,
for each  thick-time slice, 
a series of ``configurations'' for the set of indices 
$({\Rb_i,p,c,{\bf m},\nub})$ and averaging over them. There 
is clearly a great freedom in choosing the best sampling 
procedure. An efficient one
is to project exactly on the quantum numbers which 
eliminate the contributions from states lighter than the one of interest, 
while treating stochastically the indices which project out heavier 
states. One can also extract the indices 
with a distribution different from a constant one, or implement
a more involved sampling procedure. Once the estimate of the l.h.s.
of Eq.~(\ref{eq:sumrnd}) is inserted in 
the multi-level algorithm (see below), the final result is independent 
from the particular procedure implemented and from the statistical accuracy
on the thick time-slice matrix elements. The algorithm is design to be
always exact, but the variance and therefore its efficiency depends 
on details of the implementation.

\subsection{Multi-level integration scheme}
The ratio of partition functions $Z^{(\mu,\Ps,\Cs)}/Z$ can then be calculated by 
implementing the hierarchical two-level integration formula
(see Figure~\ref{fig:multiL})
\be\label{eq:bella}\displaystyle
\frac{Z^{(\mu,\Ps,\Cs)}}{Z} = \sum_{j=1}^{n_\mu} \frac{1}{Z} 
\int {\rm D}_4 [U]\, e^{-S[U]} 
\,\, {\rm P}^{(\mu,\Ps,\Cs)}_{j;m,d}\Big[T,0\Big]\; ,
\ee
where 
\be\label{eq:Psmd}
{{\rm P}}^{(\mu,\Ps,\Cs)}_{j;m,d}\Big[y_0,x_0\Big] =  
{\prod_{i=0}^{m-1}}\,
 \frac{{\rm T}^{(\mu,\Ps,\Cs)}_{jj}[U_{x_0+(i+1)\cdot d},U_{x_0 + i\cdot d}]}
     {{\rm T}[U_{x_0+(i+1)\cdot d},U_{x_0 + i\cdot d}]}
\ee
with $m\ge 1$, $y_0=(x_0 + m \cdot d)$. In each ratio on the r.h.s of 
Eq.~(\ref{eq:Psmd}) the stochastic indices are generated,
independently on each thick-time slice,
with chosen distribution $\Pi_{\R_i,p,c,{\bf m},\nubs}$.
The procedure can be generalized easily. For a three-level scheme, 
for instance, each ratio can be computed 
with a two-level scheme by exploiting the composition rule in 
Eq.~(\ref{eq:compmu}). While 
the result does not depend on the particular integration scheme 
implemented, its statistical error does. The algorithm therefore 
requires an optimization which allows one to exploit the expected 
spectral properties of the theory.
For instance if $d$ is chosen large enough, i.e. larger than 
$1/T_c$ with $T_c$ being the critical temperature, only a few of the 
physical states give a sizeable contribution to each ratio
${\rm T}^{(\mu,\Ps,\Cs)}_{jj}[U_{x_0+d},U_{x_0}]/{\rm T}[U_{x_0+d},U_{x_0}]$.
The latter is therefore expected to be of order $e^{-E^{(\mu,\Ps,\Cs)}_1\, d}$,
the magnitude of the product is of the order of 
$e^{-E^{(\mu,\Ps,\Cs)}_1\, T}$ for each configuration of the boundary fields,  
and the statistical fluctuations are reduced to this 
level~\cite{DellaMorte:2008jd,DellaMorte:2007zz}.

\subsection{Correlators of composite operators}
The formula in Eq.~(\ref{eq:factOO}) and its generalization 
to the case of multiple quantum numbers can be implemented 
by inserting the sources in a thick time-slice and then
following the same steps that lead to Eq.~(\ref{eq:bella}).
It is important to stress, however, that once all the 
thick time-slice ratios 
${\rm T}^{(\mu,\Ps,\Cs))}_{jj}[U_{x_0+d},U_{x_0}]/{\rm T}[U_{x_0+d},U_{x_0}]$ 
have been computed, only the first integral of the telescopic 
expansion in Eq.~(4.2) of Ref.~\cite{DellaMorte:2008jd} needs 
to be re-done for the thick time-slices where a source is 
inserted. The extra numerical burden is therefore negligible. 
As in the previous subsection, the statistical error 
associated to the estimate of the correlation function is 
comparable to the signal if the algorithm is 
properly optimized.

\section{Numerical results}
To demonstrate the practical feasibility of the strategy proposed 
in this paper, we have carried out a full-fledged computation of 
the mass and multiplicity of the lightest glueball with vacuum 
quantum numbers. We have simulated the $\rm{SU}(3)$ gauge 
theory at $\beta=6/g^2_0=5.7$, which corresponds to a spacing 
of $0.17$~fm if the 
reference scale $r_0=0.5$~fm is used to calibrate the 
lattice~\cite{Guagnelli:1998ud}. The spatial 
lengths of the lattices are $1.4$ and $1.7$~fm, while 
their time dimension extends up to $2$~fm.
A list of the runs, the number of configurations
generated and some details of the multi-level algorithm 
implemented is reported in Table~\ref{tab:lattices}.
\begin{table}[thb]
\vspace{0.1cm}
\begin{center}
\begin{tabular}{|cccccc|}
\hline
Lattice&$L$&$T$&$N_\mathrm{conf}$&$N_\mathrm{lev}$&$d$\\[0.125cm]
\hline
${\rm A}_1$&   8  &  4  & 50  & 2 & 4  \\[0.125cm]
${\rm A}_2$&      &  5  & 50  & 2 & 5  \\[0.125cm]
${\rm A}_3$&      &  6  & 100 & 2 & 3  \\[0.125cm]
${\rm A}_4$&      &  8  & 100 & 2 & 4  \\[0.125cm]
${\rm A}_5$&      &  12 & 50  & 3 & $\left\{3,6\right\}$ \\[0.125cm]
\hline
${\rm B}_3$&   10 &  6  & 50  & 2 & 3  \\[0.125cm]
\hline
\end{tabular}
\caption{Simulation parameters: $N_\mathrm{conf}$ is the number of configurations
of the uppermost level, $N_\mathrm{lev}$ is the number of levels and $d$ is the thickness 
of the thick time-slice used for the various levels.\label{tab:lattices}}
\end{center}
\end{table}

\begin{figure}[thb]
\begin{center}
\includegraphics[width=14.0cm]{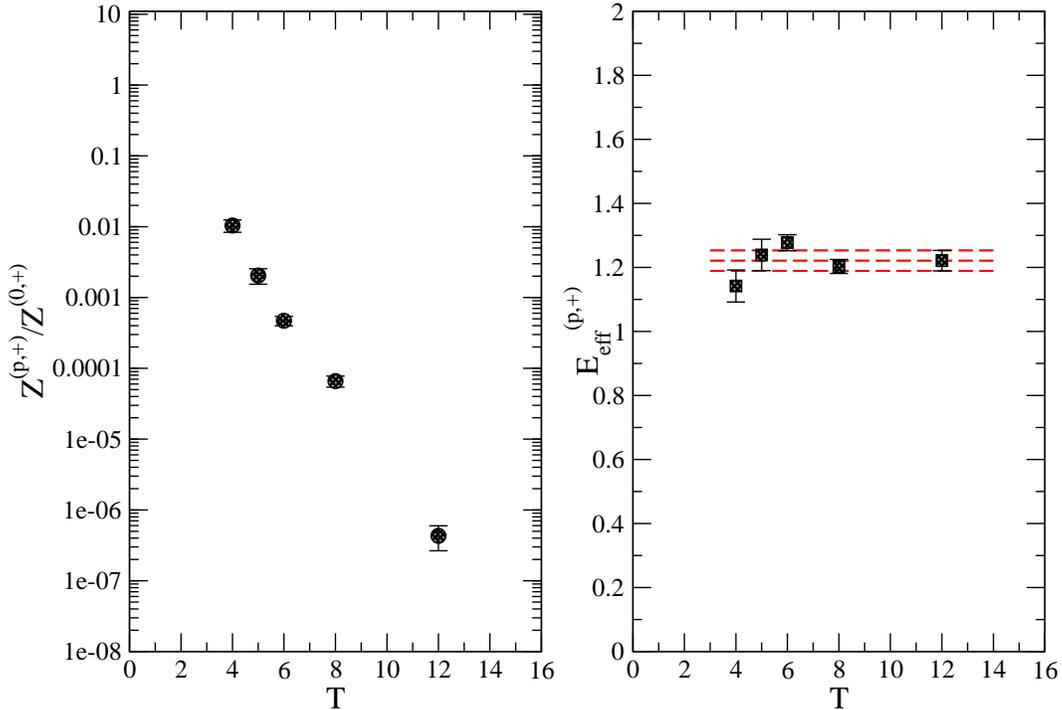}
\caption{Left panel: ratio of partition functions $Z^{({\bf p},+)}/Z^{({\bf 0},+)}$ 
with momenta ${\bf p}=[2\pi/L,0,0]$ for the ${\rm A}$ lattices. Right panel: the 
corresponding effective energy as defined in Eq.~(\ref{eq:effenergy}). The band is our 
best estimate, i.e. the one  extracted from the lattice with the longest 
time-extension.\label{fig:fig0pp}}
\end{center}
\end{figure}

The primary quantity that we have calculated numerically is the 
ratio of partition functions $Z^{({\bf p},+)}/Z$ defined 
in section \ref{sec:gmm}, with $p_{2,3}=0$, 
$p_1=(2\pi/L)\, n_1$ and $n_1=1,2$. The thick 
time-slice transfer matrix elements associated to the projector in 
Eq.~(\ref{eq:p+}) have been computed as described in 
section~\ref{sec:algo}: the sum on the group indices of
charge conjugation and translations along direction 1
has been done exactly, while the one over the 
remaining indices $(p_2,p_3,\nub)$ has been
carried out stochastically by extracting between 9 and 64 
random ``configurations'' of indices with a flat probability. 
The results are collected in Table~\ref{tab:0pp}, and those of  
the ${\rm A}$ lattices are plotted in Figure~\ref{fig:fig0pp} 
as a function of $T$ (left panel). The data show a clear
exponential decay of the ratio $Z^{({\bf p},+)}/Z^{({\bf 0},+)}$
over more than 6 orders of magnitude.

We fit the results of the ${\rm A}$ series to a single 
exponential, i.e. 
\be\displaystyle
\ln\left[\frac{Z^{({\bf p},+)}}{Z^{({\bf 0},+)}}\right] 
= A - B\, T \; ,
\label{eq:fitform}
\ee
where $A=\ln{\omega^+}$, with $\omega^+$ being the multiplicity of the state,
and $B=E_{\mbox{eff}}^{({\bf {p}},+)}$. This function fits well the last four 
points, and the best fit gives\footnote{The small value observed at $T=6$
is compatible with a $2 \sigma$ statistical fluctuation, and it is 
responsible for the value of $\chi^2/\mathrm{dof}$ a bit larger than expected.} 
$A=-0.6(4)$ and $B=1.15(6)$ ($\chi^2/\mathrm{dof}=1.5$). Group theory 
predicts the multiplicity to be an integer between $0$ and $3$. The null 
value is excluded by the data, the multiplicity 1 is within $1.5\sigma$ of the 
central value given by the fit, while 2 and 3 are $3.2\,\sigma$ and 
$4.2\sigma$ away. As a further check we also fix the multiplicity to one 
of the possible integers, and we 
obtain $\chi^2/\mathrm{dof}=1.7$, $4.4$ and $6.8$ for $\omega^+=1$, 
$2$  and $3$ respectively. We therefore conclude that the data strongly 
prefer multiplicity 1, the value expected for a singlet  
under the octahedral group. We stress once again that the computation 
of the multiplicity is new because this quantity is not accessible to  
the standard technique.
\begin{table}[thb]
\begin{center}
\begin{tabular}{|cl|clll|}
\hline
Lattice    & $Z^{({\bf 0},+)}/Z$ & $n_1$ &  $Z^{({\bf p},+)}/Z$      &$Z^{({\bf p},+)}/Z^{({\bf 0},+)}$&
$E_{\mathrm{eff}}^{({\bf {p}},+)}$\\[0.125cm]
\hline
${\rm A}_1$&        $0.158(14)$        &  $1$  &$1.6(3)\cdot 10^{-3}$  &$1.04(21)\cdot 10^{-2}$&$1.14(5)$\\[0.125cm] 
${\rm A}_2$&        $0.871(5)$         &  $1$  &$1.8(4)\cdot 10^{-3}$  &$2.0(5)\cdot 10^{-3}$  &$1.24(5)$\\[0.125cm]
${\rm A}_3$&        $0.96(4)$          &  $1$  &$4.5(7)\cdot 10^{-4}$  &$4.7(7)\cdot 10^{-4}$  &$1.277(25)$\\[0.125cm]
${\rm A}_4$&        $0.999(9)$         &  $1$  &$6.6(12)\cdot 10^{-5}$ &$6.6(12)\cdot 10^{-5}$ &$1.203(22)$\\[0.125cm]
${\rm A}_5$&        $0.963(13)$        &  $1$  &$4.1(16)\cdot 10^{-7}$ &$4.3(17)\cdot 10^{-7}$ &$1.22(3)$\\[0.125cm]
\hline
${\rm B}_3$&        $1.03(4)$          &  $1$  &$1.0(3)\cdot 10^{-3}$  &$1.0(3)\cdot 10^{-3}$  &$1.15(5)$\\[0.125cm]
           &                           &  $2$  &$0.94(25)\cdot 10^{-4}$&$0.92(25)\cdot 10^{-4}$&$1.55(5)$\\[0.125cm]
\hline
\end{tabular}
\caption{Results for ratios of partition functions with momenta ${\bf p}=[2\pi n_1/L,0,0]$. 
The effective energy $E_{\mathrm{eff}}^{({\bf {p}},+)}$ is defined as in 
Eq.~(\ref{eq:effenergy}).\label{tab:0pp}}
\end{center}
\end{table}

By imposing the multiplicity to be 1, the energy of the state can be determined
at each $T$ as 
\be
E^{({\bf p},+)}_\mathrm{eff} = -\frac{1}{T}\,
\ln\left[\frac{Z^{({\bf p},+)}}{Z^{({\bf 0},+)}}\right]\; ,
\label{eq:effenergy}
\ee
which yields the results given in Table~\ref{tab:0pp} and shown in the 
right panel of Figure~\ref{fig:fig0pp}. It is interesting to notice that 
a precision of a few percent is reached with only 50-100 configurations of the 
uppermost algorithmic level. To be on the conservative side, we take as our best 
estimate for the energy the value at $T=12$ reported in Table~\ref{tab:0pp}. The 
result of the fit to a constant of the last four data points gives $1.233(14)$, a value
compatible with our best one but with half the statistical error.

\begin{figure}[!t]
\begin{center}
\includegraphics[width=10.0cm]{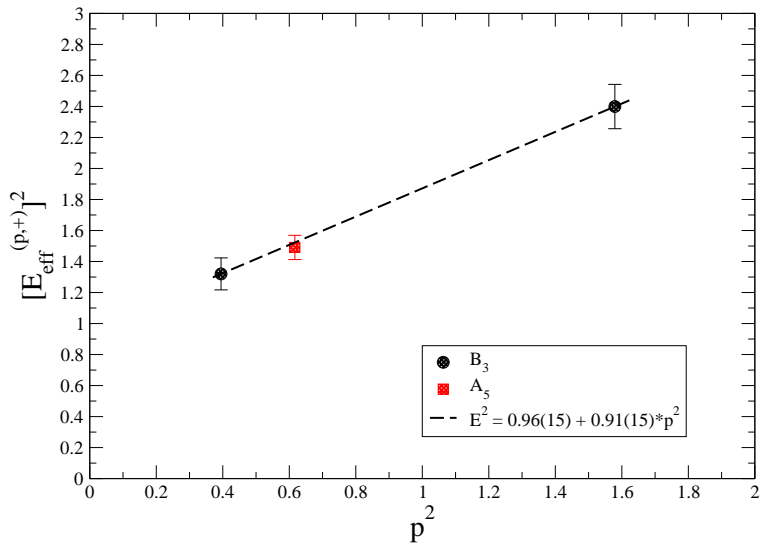}
\caption{The effective energy squared from the $A_5$ (red square) 
and the $B_3$ (black circles) lattices. The line is a linear 
interpolation of the black points (circles).
\label{fig:finite_vol}}
\end{center}
\end{figure}

Finite volume effects in the energy values are expected to be exponentially
suppressed at asymptotically large values of $L$. Lattice ${\rm B}_3$ serves 
the purpose of assessing their magnitude. It has the same lattice spacing of 
the ${\rm A}$ series but a linear extension of $L=10$. The results for 
$n_1=1,2$ are reported in 
Table~\ref{tab:0pp}, and are plotted as a function of the momentum 
squared in Figure~\ref{fig:finite_vol}.
The dashed line is a linear interpolation of the two 
black points  (circles) of the lattice ${\rm B}_3$, while the red 
point (square) is our best result for the ${\rm A}$ series. 
It is rather clear that, within our statistical precision, 
finite volume effects are not visible in our data. It is also 
interesting to notice that, even if the values of the momenta are 
rather large, the continuum dispersion relation is well reproduced 
within our statistical errors.

The glueball mass can finally be extracted, up to ${\cal O}(a^2)$ 
discretization errors, by using the  continuum dispersion relation 
\be
M^+ = \sqrt{(E_\mathrm{eff}^{({\bf p},+)})^2 - {\bf p}^2} \;,
\ee
which, in units of the lattice spacing, yields to 
\be\label{eq:best}
M^+ = 0.935 \pm 0.042\; .
\ee
This is one of the main numerical results of this paper. When converted
in physical units by using $r_0$, it gives $\mathrm{M}^+=1.08(5)$~GeV. We remark that 
the value in Eq.~(\ref{eq:best}) is fully compatible with 0.955(15), the mass 
computed with the standard method by using the 
same action at the same lattice spacing~\cite{Vaccarino:1999ku}. On the other 
hand its value in physical units is quite smaller than the continuum extrapolated 
one estimated, for instance,  in Ref.~\cite{Chen:2005mg}. This can be easily 
explained by discretization effects which still contribute to the value in 
Eq.~(\ref{eq:best})~\cite{Vaccarino:1999ku}. Removing them goes beyond 
the scope of this paper.

To identify the state of mass $M^+$ with the $0^{++}$ glueball we 
still need to verify that the lightest state in the sector orthogonal 
to the vacuum, i.e. the one selected by the projector in Eq.~(\ref{eq:vort}), 
is heavier. Since the multiplicity turns out to be one, we can 
further restrict the orthogonal sector by projecting onto the singlet 
representations of the octahedral group ${\rm A}_1$ and ${\rm A}_2$. 
The ratio of such a partition function over the standard one has 
been computed in a dedicated run at $T=6$ and $L=8$, and our 
best estimate turns out to be $2.6(26)\cdot 10^{-4}$. This result,
even if compatible with zero, puts an upper bound on this quantity 
much smaller than the one expected if a state lighter than $M^+$ were 
present. This suggests to identify $M^{+}$ with the mass of the 
$A_1^{++}$ glueball. More simulations are needed to corroborate
this result, and to have a solid numerical evidence of the 
existence of a mass gap in the theory.

\section{Conclusions}
The relative contributions to the partition function, due to states carrying 
a given set of quantum numbers associated with the exact symmetries of a 
field theory, can be expressed by ratios of path integrals with different 
boundary conditions in the time direction. From a theoretical point of view 
these are very clean quantities, which have a finite and universal continuum 
limit once the bare parameters in the action have been renormalized. From an 
algorithmic point of view, the composition properties of the projectors 
can be exploited to implement a hierarchical multi-level integration procedure 
which solves the problem of the exponential (in time) degradation of 
the signal-to-noise ratio. 

The numerical study presented in this paper demonstrates that 
glueball masses can be extracted from these observables (for the SU(3) 
Yang--Mills theory) with the present generation of computers. We have been
able to follow the exponential decay of a ratio of partition functions
dominated by the lightest glueball with vacuum quantum numbers
for more than 6 orders of magnitude. A fit to a single exponential in 
the time range $0.85$~--~$2$~fm has allowed us to determine unambiguously 
the multiplicity of the lightest state for the first time. The 
mass has then been extracted with a precision of few percent at a time 
distance of $2$~fm, where the contamination from excited states is 
negligible. 

The ideas presented here can also lead to a solid quantitative evidence 
of the existence of a mass gap in the theory. To this aim a continuum limit 
extrapolation of the numerical results is mandatory. At present, however, 
the scaling of the algorithm with the square of the number of spatial points, 
together with the limited numerical resources at our disposal, prevents 
us to simulate the theory at 
small enough lattice spacings for such a limit to be reliably taken. Another 
intriguing application is the computation 
of the thermodynamic potentials at non-zero temperature as suggested in 
Ref.~\cite{Giusti:2010bb}. No vacuum subtraction or renormalization constant 
is required, and the method can 
be applied at arbitrary high and low temperatures. The ratios introduced 
here may also turn out to 
be useful in the study of QCD-like theories, where isolating the contributions 
from single states in correlation functions may not be easy. 

\section*{Acknowledgments}
We thank M. L\"uscher, H. B. Meyer and R. Sommer for interesting 
discussions. The simulations were performed on PC clusters at CERN,
CILEA, at the Swiss National Supercomputing Centre (CSCS) and 
at the J\"ulich Supercomputing Centre (JSC). We thankfully acknowledge 
the computer resources and technical support provided by all these 
institutions and their technical staff. 

\appendix 

\section{Lagrangian of the Wilson action \label{app:appA}}
The Lagrangian $L$ which enters the definition of the transfer 
matrix elements in Eq.~(\ref{eq:TW}) is given by 
\be
L\Big[U_{x_0+1},U_{x_0}\Big] =  K\Big[U_{x_0+1},U_{x_0}\Big]
+  \frac{1}{2}W\Big[U_{x_0+1}\Big] + \frac{1}{2}W\Big[U_{x_0}\Big]\; ,
\ee
where the kinetic and the potential contributions are defined as 
\be
K\Big[U_{x_0+1},U_{x_0}\Big] = \beta \sum_{{\bf x},k} 
\left[ 1 - \frac{1}{3}{\rm Re}\Tr\left\{ 
U_{k}(x_0+1,{\bf x}) U^\dagger_{k}(x_0,{\bf x}) 
\right\}\right]
\ee
and 
\be
W\Big[U_{x_0}\Big] = \frac{\beta}{2} \sum_{{\bf x}} \sum_{k,l} 
\left[1 - \frac{1}{3}{\rm Re}\Tr\Big\{U_{kl}(x_0,{\bf x})\Big\}\right]\;  
\ee
respectively.

\section{Irreducible representations of the octahedral group \label{app:appB}}
In this appendix we report our conventions for the octahedral group and its
irreducible representations. The group has 24 elements $\Rb_i$, 
$3\times 3$ orthogonal matrices ($\Rb^T_i \Rb_i=\id$), grouped in five conjugacy classes. 

\noindent First class $(E)$:
\be\label{eq:Rmu1}
\Rb_1 = 
\left(\begin{array}{ccc}
1 & 0 & 0\\
0 & 1 & 0\\
0 & 0 & 1
\end{array}\right)\; .
\ee
\noindent Second class $(C_3)$:
\bea\label{eq:Rmu2}
\Rb_2 & = & 
\left(\begin{array}{ccc}
0 &  1 & 0\\
0 &  0 &-1\\
-1&  0 & 0
\end{array}\right)
\; , \quad
\Rb_3 = 
\left(\begin{array}{ccc}
0 &-1 & 0\\
0 & 0 &-1\\
1 & 0 & 0
\end{array}\right)
\; , \quad
\Rb_4 = 
\left(\begin{array}{ccc}
0 &-1 & 0\\
0 & 0 & 1\\
-1& 0 & 0
\end{array}\right)\nonumber\\[0.5cm]
\Rb_5 & = & 
\left(\begin{array}{ccc}
\;\; 0 &\;\; 1 & \; 0\;\\
\;\; 0 &\;\; 0 & \; 1\;\\
\;\; 1 &\;\; 0 & \; 0\;
\end{array}\right)
\; , \quad
\Rb_6 = 
\left(\begin{array}{ccc}
0 & 0 &-1\\
1 & 0 & 0\\
0 &-1 & 0
\end{array}\right)
\; , \quad
\Rb_7 = 
\left(\begin{array}{ccc}
0 & 0 & 1\\
-1& 0 & 0\\
 0&-1 & 0
\end{array}\right)\nonumber\\[0.5cm]
\Rb_8 & = & 
\left(\begin{array}{ccc}
0 & 0 &-1\\
-1& 0 & 0\\
0 & 1 & 0
\end{array}\right)
\; , \quad\;
\Rb_9 = 
\left(\begin{array}{ccc}
\;0\; &\; 0\; &\; 1\\
\;1\; &\; 0\; &\; 0\\
\;0\; &\; 1\; &\; 0
\end{array}\right)\; .
\eea

\noindent Third class $(C_2)$:
\be\label{eq:Rmu3}
\Rb_{10} =  
\left(\begin{array}{ccc}
1 & 0 & 0\\
0 &-1 & 0\\
0 & 0 &-1
\end{array}\right)
\; , \;\;
\Rb_{11} = 
\left(\begin{array}{ccc}
-1& 0 & 0\\
0 & 1 & 0\\
0 & 0 &-1
\end{array}\right)
\; , \;\;
\Rb_{12} = 
\left(\begin{array}{ccc}
-1& 0 & 0\\
0 &-1 & 0\\
0 & 0 & 1
\end{array}\right)\; .
\ee

\noindent Fourth class $(C_4)$:
\bea\label{eq:Rmu4}
\Rb_{13} & = & 
\left(\begin{array}{ccc}
1 & 0 & 0\\
0 & 0 & 1\\
0 &-1 & 0
\end{array}\right)
\; , \quad
\Rb_{14} = 
\left(\begin{array}{ccc}
0 & 0 &-1\\
0 & 1 & 0\\
1 & 0 & 0
\end{array}\right)
\; , \quad
\Rb_{15} = 
\left(\begin{array}{ccc}
0 & 1 & 0\\
-1& 0 & 0\\
0 & 0 & 1
\end{array}\right)\; \\[0.5cm]
\Rb_{16} & = & 
\left(\begin{array}{ccc}
1 & 0 & 0\\
0 & 0 &-1\\
0 & 1 & 0
\end{array}\right)
\; , \quad
\Rb_{17} = 
\left(\begin{array}{ccc}
 0& 0 & 1\\
0 & 1 & 0\\
-1& 0 & 0
\end{array}\right)
\; , \quad
\Rb_{18} = 
\left(\begin{array}{ccc}
0 &-1 & 0\\
1 & 0 & 0\\
0 & 0 & 1
\end{array}\right)\; .\nonumber
\eea

\noindent Fifth class $(C_2')$:
\bea\label{eq:Rmu5}
\Rb_{19} & = & 
\left(\begin{array}{ccc}
0 & 1 & 0\\
1 & 0 & 0\\
0 & 0 &-1
\end{array}\right)
\; , \quad
\Rb_{20} = 
\left(\begin{array}{ccc}
0 &-1 & 0\\
-1& 0 & 0\\
0 & 0 &-1
\end{array}\right)
\; , \quad
\Rb_{21} = 
\left(\begin{array}{ccc}
0 & 0 & 1\\
0 &-1 & 0\\
1 & 0 & 0
\end{array}\right)\;\;\; \\[0.5cm]
\Rb_{22} & = & 
\left(\begin{array}{ccc}
0 & 0 &-1\\
0 &-1 & 0\\
-1& 0 & 0
\end{array}\right)
\; , \quad
\Rb_{23} = 
\left(\begin{array}{ccc}
-1& 0 & 0\\
0 & 0 & 1\\
0 & 1 & 0
\end{array}\right)
\; , \quad
\Rb_{24} = 
\left(\begin{array}{ccc}
-1 & 0 & 0\\
 0& 0 &-1\\
 0&-1 & 0
\end{array}\right)\; .\nonumber
\eea

\noindent The group has 5 inequivalent irreducible
representations.

\noindent Singlet $A_1$:
\be
\Gamma(\Rb_i) = 1 \; \qquad i=1,\dots,24\; .
\ee

\noindent Singlet $A_2$:
\[
\Gamma(\Rb_i) = 1 \; \qquad i=1,\dots,12
\]
\be
\Gamma(\Rb_i) = -1 \; \qquad i=13,\dots,24\; .
\ee

\noindent Doublet $E$: 
\[
\Gamma(\Rb_1)=\Gamma(\Rb_{10})=\Gamma(\Rb_{11})=\Gamma(\Rb_{12})= \id\; ,
\]

\[
\Gamma(\Rb_2)=\Gamma(\Rb_{3})=\Gamma(\Rb_{4})=\Gamma(\Rb_{5})= 
-\cos{\left(\frac{\pi}{3}\right)}\, \id - i \sin{\left(\frac{\pi}{3}\right)}\, \sigma_2\; , 
\]

\be
\Gamma(\Rb_6)=\Gamma(\Rb_{7})=\Gamma(\Rb_{8})=\Gamma(\Rb_{9})= 
-\cos{\left(\frac{\pi}{3}\right)}\, \id + i \sin{\left(\frac{\pi}{3}\right)}\, \sigma_2 \; ,
\ee

\[
\Gamma(\Rb_{13})=\Gamma(\Rb_{16})=\Gamma(\Rb_{23})=\Gamma(\Rb_{24})= 
\cos{\left(\frac{\pi}{3}\right)}\, \sigma_3 - \sin{\left(\frac{\pi}{3}\right)}\, \sigma_1 \; ,
\]

\[
\Gamma(\Rb_{14})=\Gamma(\Rb_{17})=\Gamma(\Rb_{21})=\Gamma(\Rb_{22})= 
\cos{\left(\frac{\pi}{3}\right)}\, \sigma_3 + \sin{\left(\frac{\pi}{3}\right)}\, \sigma_1 \; ,
\]

\[
\Gamma(\Rb_{15})=\Gamma(\Rb_{18})=\Gamma(\Rb_{19})=\Gamma(\Rb_{20})= - \sigma_3 \; ,
\]
where $\id$ is the $2\times 2$ identity matrix and $\sigma_i$
are the Pauli matrices. 

\noindent Defining triplet $T_1$:
\be
\Gamma(\Rb_i) = \Rb_i \; \qquad i=1,\dots,24\; .
\ee
\noindent Triplet $T_2$:
\bea
\Gamma(\Rb_i) & = & \Rb_i \; \qquad \;\;\; i=1,\dots,12\nonumber\\
\Gamma(\Rb_i) & = & - \Rb_i \; \qquad i=13,\dots,24\; .
\eea
\begin{table}[!t]
\begin{center}
\begin{tabular}{cc|ccccc}
           &$\mu$&$E$&$8\, C_3$&$3\, C_2$&$6\, C_4$&$6\, C_2'$\\[0.125cm]
\hline
${\rm A}_1$&1& 1 & 1 & 1 & 1 & 1 \\[0.125cm]
${\rm A}_2$&2& 1 & 1 & 1 &-1 &-1 \\[0.125cm]
${\rm E}$  &3& 2 &-1 & 2 & 0 & 0 \\[0.125cm]
${\rm T}_1$&4& 3 & 0 &-1 & 1 &-1 \\[0.125cm]
${\rm T}_2$&5& 3 & 0 &-1 &-1 & 1 \\
\end{tabular}
\caption{Table of characters for the octahedral group\label{tab:char}}
\end{center}
\end{table}
The characters of the various representations are reported in 
Table~\ref{tab:char}.

\bibliographystyle{h-elsevier}   
\bibliography{lattice}        
\end{document}